\documentclass[journal]{IEEEtran}
\usepackage{cite}
\ifCLASSINFOpdf
  \usepackage[pdftex]{graphicx}
  \DeclareGraphicsExtensions{.pdf,.jpeg,.png}
\else
  \usepackage[dvips]{graphicx}
  \DeclareGraphicsExtensions{.eps}
\fi
\usepackage{amsmath}
\begin{document}
\title{Simulation of Circular Cylindrical Metasurfaces using GSTC-MoM}


\author{Srikumar~Sandeep,~\IEEEmembership{Member,~IEEE,}
        and~Shao~Ying~Huang,~\IEEEmembership{Member,~IEEE}% <-this % stops a space
%\thanks{M. Shell was with the Department
%of Electrical and Computer Engineering, Georgia Institute of Technology, Atlanta,
%\thanks{J. Doe and J. Doe are with .}% <-this % stops a space
\thanks{S.\ Sandeep (email: sandeepsrikumar2013@gmail.com) and S. Y. \ Huang (email: huangshaoying@sutd.edu.sg) are with Singapore University of Technology and Design, Singapore}
}
%
%\thanks{Manuscript received April 19, 2016; revised August 26, 2016.}}


\markboth{JMMCT}%,%~Vol.~14, No.~8, August~2016}%
{Shell \MakeLowercase{\textit{et al.}}: Bare Demo of IEEEtran.cls for IEEE Journals}



\maketitle

\begin{abstract}
A modeling of circular cylindrical metasurfaces using Method of Moments (MoM) based on Generalized Sheet Transition Conditions (GSTCs) is presented. GSTCs are used to link the integral equations for fields on the inner and outer contour of the cylindrical metasurface. The GSTC-MoM is validated by a case of an anisotropic, gyrotropic metasurface capable of two field transformations. The formulations presented here can be used as a platform for deriving GSTC-MoM for 3D spherical and conformal metasurfaces. 
\end{abstract}


% \fi
\IEEEpeerreviewmaketitle

\section{Introduction}
\IEEEPARstart{M}{etasurfaces} are deeply subwavelength surfaces which can manipulate electromagnetic waves in a desired manner \cite{HollawayMetasurfaceReview}. Essentially, these are field transformers which are constructed by arrangement of subwavelength scatterers in a host medium. Metasurfaces have practical advantages over bulk metamaterials including easier fabrication, lower loss and less weight \cite{MetamaterialsSolymar}. Even though they have similarities with frequency selective surfaces \cite{MunkFSS}, the design possibilities offered by metasurfaces are much broader. Metasurface applications include polarization transformation \cite{PfeifferGrbicPolControl}, 2D waveguides \cite{SpatialWaveguideMetasurface}, radiation pressure control \cite{AchouriCalozSolarSail}, generalized refraction \cite{NYuLightProp}, broadband absorbers \cite{MetasurfaceBroadbandAbsorber}, flat optical components \cite{FlatOpticsYu}, LED efficiency enhancers \cite{LEDEnhancerMS}, spatial isolators \cite{SpatialIsolatorMS} etc. Review of metasurface and its applications can be found in \cite{RecentDevelopmentsMSAchouri,MetasurfaceReviewHTChen,MetasurfaceReviewGlybovski}.

Metasurfaces achieve its functionality by creating a spatio-temporal electromagnetic discontinuity. Mathematically, the discontinuity can be expressed by Generalized Sheet Transition Conditions (GSTC) which relates the electric and magnetic field discontinuities to the electric and magnetic surface polarization current densities \cite{IdemenBook,achouri2015general}.  At present, commercial electromagnetic simulation softwares can model several boundary conditions, such as perfect electric conductor (PEC), perfect magnetic conductor (PMC), periodic boundary condition (PBC), standard impedance boundary condition (SIBC), radiation boundary condition (RBC), and perfectly matched layer (PML). However, no commercial CAD tools have yet incorporated the modeling of GSTCs. Therefore, it is important to develop numerical modeling of GSTCs for analysis and synthesis of metasurfaces. The modeling of GSTCs in the Finite Difference Frequency Domain (FDFD) method was reported in \cite{GSTCFDFD}.This work was extended to handle a more general dispersive, time varying metasurface using a Finite Difference Time Domain (FDTD)-GSTC formulation in \cite{GSTCFDTD,GSTCPLRCFDTD}. Modeling of GSTCs in the finite element method (FEM), which is one of the more widely used numerical methods to simulate practical problems was described in \cite{GSTCFEM}. An Integral Equation (IE) solution to planar, time varying metasurface was described in \cite{GSTCSDIE}. A review of computational electromagnetic methods applied to metasurface analysis can be found in \cite{ComputationalAnalysisMS}.

The vast majority of metasurfaces reported to date are planar. Other canonical shapes such as cylindrical metasurfaces and spherical metasurfaces \cite{SphericalMS} are now being studied. It is expected that conformal metasurfaces (metasurfaces of irregular shape) will be a subject of active research \cite{SphericalMS}. Some of applications of cylindrical metasurfaces include leaky wave antennas \cite{CircumModCylMS,LWACylindricalMS}, radiation pattern control \cite{RadPatternControlCylMS} and cloaking \cite{CylindricalMSCloaking}. The goal of this paper is to analyze cylindrical metasurface scattering problem using Method of Moments (MoM). The formulation presented here paves the way for IE-MoM based solution for spherical and conformal metasurfaces \cite{SphericalMS}. IE-MoM have the advantage over FEM and FDTD of not requiring to mesh the entire problem space. This will provide tremendous computational capability for electrically large problems involving metasurfaces. It should be noted that physical metasurfaces have a finite subwavelength thickness. Simulating such structures directly would result in very dense meshes around the metasurfaces and hence compromise the simulation efficiency. By replacing a physical metasurface by an equivalent GSTC, the burden of mesh generation can be reduced significantly and the simulation efficiency can be enhanced considerably. This is particularly important in simulation scenarios where multiple metasurfaces are involved or when repetitive simulations are required for physical metasurfe design and optimization \cite{achouri2015general,ComputationalAnalysisMS}.  

The organization of the paper is as follows. Section II recalls the GSTC metasurface synthesis equations. This is followed by a summary of 2D integral equations in Section III. Section IV shows the derivation of GSTC-MoM for 2D cylindrical problems. A numerical validation of the derived formulation is shown in Section V. Conclusions are provided in Section VII. 

 

\section{Cylindrical metasurface synthesis equations}
Metasurface synthesis equations for planar metasurfaces and spherical metasurfaces are described in \cite{achouri2015general} and \cite{SphericalMS} respectively. Similar to \cite{achouri2015general} and \cite{SphericalMS}, the normal polarization current densities are ignored.  Following the bianisotropic susceptibility-GSTC approach, the metasurface synthesis equations for a cylindrical metasurface of radius $a$ with its axis along $z$ direction are given by
\begin{subequations}
\label{GSTCEqn}
\begin{align}
\big[\hat{\rho} \times \Delta \bar{H} &= j\omega \bar{P}_{\mathrm{s},||} \big]\big|_{\rho = a} \\
\big[\hat{\rho} \times \Delta \bar{E} &= -j\omega \mu_{0} \bar{M}_{\mathrm{s},||} \big]\big|_{\rho = a}
\end{align}
\end{subequations}
where $\bar{P}_{\mathrm{s},||},\bar{M}_{\mathrm{s},||}$ are the transverse electric and magnetic surface polarization densities. The medium internal and external to the metasurface cylinder is free space. $\Delta \Psi = \Psi^{+} - \Psi^{-}$ denote the jump discontinuity of field component $\Psi$.  A time harmonic dependance of $e^{j\omega t}$ is assumed.
 \begin{subequations}
\label{PolDensityEqn}
\begin{align}
\bar{P}_{\mathrm{s},||} &= \epsilon_{o} \bar{\bar{\chi}}_{\mathrm{ee}} \bar{E}_{\mathrm{\mathrm{av}}} + \sqrt{\mu_{0} \epsilon_{0}}
\bar{\bar{\chi}}_{\mathrm{em}} \bar{H}_{\mathrm{\mathrm{av}}} \\
\bar{M}_{\mathrm{s},||} &= \sqrt{\frac{\epsilon_{o}}{\mu_{0}}} \bar{\bar{\chi}}_{\mathrm{me}} \bar{E}_{\mathrm{av}} + 
\bar{\bar{\chi}}_{\mathrm{mm}} \bar{H}_{\mathrm{av}} 
\end{align}
\end{subequations}
where $\overline{\overline{\chi}}_{\mathrm{ee}},\overline{\overline{\chi}}_{\mathrm{mm}},\overline{\overline{\chi}}_{\mathrm{em}}$, and $\overline{\overline{\chi}}_{\mathrm{me}}$ are the electric/magnetic (first e/m subscripts) susceptibility tensors describing the response to the electric/magnetic (second e/m subscripts) excitations, and the subscript ``av'' denotes the average of the fields on both sides of the metasurface, $\vec{\psi}_{\mathrm{av}} = [(\vec{\psi}^{\mathrm{inc}} + \vec{\psi}^{\mathrm{ref}}) + \vec{\psi}^{\mathrm{tr}}]/2$. Substituting (\ref{PolDensityEqn}) into (\ref{GSTCEqn}) results in the following metasurface synthesis equations:
\begin{subequations}
\label{CylMSSynthesisEqns}
\begin{equation} \label{GSTCexpandedeqns1}
\begin{split}
\begin{bmatrix}
-\Delta H_{z} \\
\Delta H_{\phi} 
\end{bmatrix}
 =  j\omega\epsilon_{0}
\begin{bmatrix}
\chi_{\mathrm{ee}}^{\phi \phi} & \chi_{\mathrm{ee}}^{\phi z} \\
\chi_{\mathrm{ee}}^{z \phi} & \chi_{\mathrm{ee}}^{zz}
\end{bmatrix}
\begin{bmatrix}
E_{\phi,\mathrm{av}} \\
E_{z,\mathrm{av}}
\end{bmatrix}
  + j\omega\sqrt{\mu_{0} \epsilon_{0}} \\
\begin{bmatrix}
\chi_{\mathrm{em}}^{\phi \phi} & \chi_{\mathrm{em}}^{\phi z} \\
\chi_{\mathrm{em}}^{z \phi} & \chi_{\mathrm{em}}^{zz}
\end{bmatrix}
\begin{bmatrix}
H_{\phi,\mathrm{av}} \\
H_{z,\mathrm{av}}
\end{bmatrix} 
\end{split}
\end{equation} 

\begin{equation}
\begin{split}
\label{GSTCexpandedeqns2}
\begin{bmatrix}
\Delta E_{z} \\
-\Delta E_{\phi}
\end{bmatrix}
  =  j\omega \sqrt{\mu_{0} \epsilon_{0}}
\begin{bmatrix}
\chi_{\mathrm{me}}^{\phi \phi} & \chi_{\mathrm{me}}^{\phi z} \\
\chi_{\mathrm{me}}^{z \phi} & \chi_{\mathrm{me}}^{zz}
\end{bmatrix}
\begin{bmatrix}
E_{\phi,\mathrm{av}} \\
E_{z,\mathrm{av}}
\end{bmatrix}
  + j\omega \mu_{0} \\
\begin{bmatrix}
\chi_{\mathrm{mm}}^{\phi \phi} & \chi_{\mathrm{mm}}^{\phi z} \\
\chi_{\mathrm{mm}}^{z \phi} & \chi_{\mathrm{mm}}^{zz}
\end{bmatrix}
\begin{bmatrix}
H_{\phi,\mathrm{av}} \\
H_{z,\mathrm{av}}
\end{bmatrix}
\end{split}
\end{equation}
\end{subequations}
which are applicable for a general bianisotropic metasurface. Through out this work, we have assumed a monoanisotropic metasurface, i.e. $\overline{\overline{\chi}}_{\mathrm{em}} = \overline{\overline{\chi}}_{\mathrm{me}} = 0$. In such a case, the metasurface synthesis equations simplifies to 
\begin{subequations}
\label{CylMSSynthesisEqnsMonoAniso}
\begin{equation} \label{GSTCexpandedeqns1MonoAniso}
\begin{split}
\begin{bmatrix}
-\Delta H_{z} \\
\Delta H_{\phi} 
\end{bmatrix}
 =  j\omega\epsilon_{0}
\begin{bmatrix}
\chi_{\mathrm{ee}}^{\phi \phi} & \chi_{\mathrm{ee}}^{\phi z} \\
\chi_{\mathrm{ee}}^{z \phi} & \chi_{\mathrm{ee}}^{zz}
\end{bmatrix}
\begin{bmatrix}
E_{\phi,\mathrm{av}} \\
E_{z,\mathrm{av}}
\end{bmatrix} 
\end{split}
\end{equation} 
\begin{equation}
\begin{split}
\label{GSTCexpandedeqns2MonoAniso}
\begin{bmatrix}
\Delta E_{z} \\
-\Delta E_{\phi}
\end{bmatrix}
  =  
 j\omega \mu_{0} 
\begin{bmatrix}
\chi_{\mathrm{mm}}^{\phi \phi} & \chi_{\mathrm{mm}}^{\phi z} \\
\chi_{\mathrm{mm}}^{z \phi} & \chi_{\mathrm{mm}}^{zz}
\end{bmatrix}
\begin{bmatrix}
H_{\phi,\mathrm{av}} \\
H_{z,\mathrm{av}}
\end{bmatrix}
\end{split}
\end{equation}
\end{subequations}

\section{2D Integral Equations}
Consider a closed circular contour $\Gamma$ of radius $a$ in the $xy$ plane. This contour represents the cylindrical GSTC surface. The domain inside $\Gamma$ is denoted by $\Omega_{2}$ and the domain outside $\Gamma$ is denoted by $\Omega_{1}$. Then for TM polarization ($E_{z},H_{\phi}$), the IEs for domains $\Omega_{1}$ and $\Omega_{2}$ are given by the following equations
\begin{subequations}
\label{TMIEEqn}
\begin{equation*}
\begin{split}
E_{z1}^{\mathrm{inc}}(\bar{\rho}) + \oint_{\Gamma} \Big[E_{z1}(\bar{\rho}^{'}) \frac{\partial G_{0}(\bar{\rho};\bar{\rho}^{'})}{\partial n^{'}}   -  \\ j\omega \mu_{0} H_{\phi1}(\bar{\rho}^{'}) G_{0}(\bar{\rho};\bar{\rho}^{'}) \Big]\ d\Gamma^{'} = 
\end{split}
\end{equation*}
\begin{equation}
=\left\{
\begin{array}{ll}
E_{z1}(\bar{\rho})\ \ \ \ \ \ ; \ \bar{\rho} \in \Omega_{1} \\
0.5	E_{z1}(\bar{\rho})\ \ ; \ \bar{\rho} \in \Gamma \\						
0 \ \ \ \ \ \ \ \ \ \ \ \ \ ; \ \bar{\rho} \in \Omega_{2}
\end{array}
\right.
\end{equation}
\begin{equation*}
\begin{split}
E_{z2}^{\mathrm{inc}}(\bar{\rho}) - \oint_{\Gamma} \Big[E_{z2}(\bar{\rho}^{'}) \frac{\partial G_{0}(\bar{\rho};\bar{\rho}^{'})}{\partial n^{'}}   -  \\ j\omega \mu_{0} H_{\phi2}(\bar{\rho}^{'}) G_{0}(\bar{\rho};\bar{\rho}^{'}) \Big]\ d\Gamma^{'} = 
\end{split}
\end{equation*}
\begin{equation}
=\left\{
\begin{array}{ll}
0\ \ \ \ \ \ \ \ \ \ \ \ \ ; \ \bar{\rho} \in \Omega_{1} \\
0.5	E_{z2}(\bar{\rho})\ \ ; \ \bar{\rho} \in \Gamma \\						
E_{z2}(\bar{\rho}) \ \ \ \ \ \  ; \ \bar{\rho} \in \Omega_{2}
\end{array} 
\right.
\end{equation}
\end{subequations}
where $G_{0}(\bar{\rho};\bar{\rho}^{'}) = \frac{1}{4j} H_{0}^{(2)}(k_{0}|\bar{\rho}-\bar{\rho}^{'}|)$ is the 2D free space Green's function \cite{JinCEMBook}. $E_{z1}(\bar{\rho}),H_{\phi1}(\bar{\rho})$ are the fields in domain $\Omega_{1}$ and $E_{z2}(\bar{\rho}),H_{\phi2}(\bar{\rho})$ are the fields in domain $\Omega_{2}$. $E_{z1}^{\mathrm{inc}}(\bar{\rho})$ is the incident electric field due to the sources in $\Omega_{1}$ and $E_{z2}^{\mathrm{inc}}(\bar{\rho})$ is the incident electric field due to sources in $\Omega_{2}$. 

Similarly for TE polarization ($E_{\phi},H_{z}$) the IEs for domain $\Omega_{1}$ and $\Omega_{2}$ are given by
\begin{subequations}
\label{TEIEEqn}
\begin{equation*}
\begin{split}
H_{z1}^{\mathrm{inc}}(\bar{\rho}) + \oint_{\Gamma} \Big[H_{z1}(\bar{\rho}^{'}) \frac{\partial G_{0}(\bar{\rho};\bar{\rho}^{'})}{\partial n^{'}}   + \\ j\omega \epsilon_{0} E_{\phi1}(\bar{\rho}^{'}) G_{0}(\bar{\rho};\bar{\rho}^{'}) \Big]\ d\Gamma^{'} = 
\end{split}
\end{equation*}
\begin{equation}
=\left\{
\begin{array}{ll}
H_{z1}(\bar{\rho})\ \ \ \ \ \ ; \ \bar{\rho} \in \Omega_{1} \\
0.5	H_{z1}(\bar{\rho})\ \ ; \ \bar{\rho} \in \Gamma \\						
0 \ \ \ \ \ \ \ \ \ \ \ \ \ ; \ \bar{\rho} \in \Omega_{2}
\end{array}
\right.
\end{equation}
\begin{equation*}
\begin{split}
H_{z2}^{\mathrm{inc}}(\bar{\rho}) - \oint_{\Gamma} \Big[H_{z2}(\bar{\rho}^{'}) \frac{\partial G_{0}(\bar{\rho};\bar{\rho}^{'})}{\partial n^{'}}   +  \\ j\omega \epsilon_{0} E_{\phi2}(\bar{\rho}^{'}) G_{0}(\bar{\rho};\bar{\rho}^{'}) \Big]\ d\Gamma^{'} = 
\end{split}
\end{equation*}
\begin{equation}
=\left\{
\begin{array}{ll}
0\ \ \ \ \ \ \ \ \ \ \ \ \ ; \ \bar{\rho} \in \Omega_{1} \\
0.5	H_{z2}(\bar{\rho})\ \ ; \ \bar{\rho} \in \Gamma \\						
H_{z2}(\bar{\rho}) \ \ \ \ \ \  ; \ \bar{\rho} \in \Omega_{2}
\end{array} 
\right.
\end{equation}
\end{subequations}
where  $E_{\phi 1}(\bar{\rho}),H_{z1}(\bar{\rho})$ are the fields in domain $\Omega_{1}$ and $E_{\phi 2}(\bar{\rho}),H_{z2}(\bar{\rho})$ are the fields in domain $\Omega_{2}$. $H_{z1}^{\mathrm{inc}}(\bar{\rho})$ is the incident magnetic field due to sources in domain 1 and $H_{z2}^{\mathrm{inc}}(\bar{\rho})$ is the incident electric field due to sources in domain 2.

It should be noted for transverse field components we have ignored $\hat{\rho}$ component. This is due to the fact that we have ignored normal susceptibility components. Both TM and TE polarizations should be considered in domains $\Omega_{1}$ and $\Omega_{2}$ because the metasurface in general can be gyrotropic. There are 8 unknowns in the above equations. They are the fields just outside the GSTC surface: $E_{z 1}(\bar{\rho}^{'}),H_{\phi 1}(\bar{\rho}^{'}),E_{\phi 1}(\bar{\rho}^{'}),H_{z 1}(\bar{\rho}^{'})$ and fields just inside the GSTC surface: $E_{z 2}(\bar{\rho}^{'}),H_{\phi 2}(\bar{\rho}^{'}),E_{\phi 2}(\bar{\rho}^{'}),H_{z 2}(\bar{\rho}^{'})$, where $\bar{\rho}^{'} \in \Gamma$. Once these field components are known, field anywhere can be obtained by (\ref{TMIEEqn}) and (\ref{TEIEEqn}).

\section{GSTC-MoM formulation}
In this section GSTC-MoM is derived by combining the IEs from section III with GSTC synthesis equations from section II.The 8 unknown quantities given by 
\begin{subequations}
\label{F1F2_eqn}
\begin{equation}
\mathbf{F_{2}(\bar{\rho}^{'})} = \Big[E_{z 2}(\bar{\rho}^{'})\ H_{\phi 2}(\bar{\rho}^{'})\ E_{\phi 2}(\bar{\rho}^{'})\ H_{z 2}(\bar{\rho}^{'})\Big]^ {T}
\end{equation}
\begin{equation}
\mathbf{F_{1}(\bar{\rho}^{'})} = \Big[E_{z 1}(\bar{\rho}^{'})\ H_{\phi 1}(\bar{\rho}^{'})\ E_{\phi 1}(\bar{\rho}^{'})\ H_{z 1}(\bar{\rho}^{'})\Big]^ {T}
\end{equation}
\end{subequations}
In (\ref{F1F2_eqn}), $\mathbf{F_{2}}$ and $\mathbf{F_{1}}$ are the fields on the inner and outer contours of the circular cylindrical metasurface. These 
are solved by using 4 IEs and metasurface synthesis equations. The IEs are obtained from $\rho \in \Gamma$ condition in equations (\ref{TMIEEqn}) and (\ref{TEIEEqn}).
\begin{subequations}
\begin{equation}
\label{IE1}
\begin{split}
\frac{E_{z1}(\bar{\rho})}{2} - \oint_{\Gamma} \Big[E_{z1}(\bar{\rho}^{'}) \frac{\partial G_{0}(\bar{\rho};\bar{\rho}^{'})}{\partial n^{'}}   -  j\omega \mu_{0} \\H_{\phi1}(\bar{\rho}^{'}) G_{0}(\bar{\rho};\bar{\rho}^{'}) \Big]\ d\Gamma^{'} = E_{z1}^{\mathrm{inc}}(\bar{\rho})\ ; \ \bar{\rho} \in \Gamma
\end{split}
\end{equation}
\begin{equation}
\label{IE2}
\begin{split}
\frac{E_{z2}(\bar{\rho})}{2} + \oint_{\Gamma} \Big[E_{z2}(\bar{\rho}^{'}) \frac{\partial G_{0}(\bar{\rho};\bar{\rho}^{'})}{\partial n^{'}}   -   j\omega \mu_{0} \\ H_{\phi2}(\bar{\rho}^{'}) G_{0}(\bar{\rho};\bar{\rho}^{'}) \Big]\ d\Gamma^{'} = E_{z2}^{\mathrm{inc}}(\bar{\rho})\ ; \ \bar{\rho} \in \Gamma
\end{split}
\end{equation}
\begin{equation}
\label{IE3}
\begin{split}
\frac{H_{z1}(\bar{\rho})}{2} - \oint_{\Gamma} \Big[H_{z1}(\bar{\rho}^{'}) \frac{\partial G_{0}(\bar{\rho};\bar{\rho}^{'})}{\partial n^{'}}   +   j\omega \epsilon_{0} \\ E_{\phi1}(\bar{\rho}^{'}) G_{0}(\bar{\rho};\bar{\rho}^{'}) \Big]\ d\Gamma^{'} = H_{z1}^{\mathrm{inc}}(\bar{\rho})\ ; \ \bar{\rho} \in \Gamma
\end{split}
\end{equation}
\begin{equation}
\label{IE4}
\begin{split}
\frac{H_{z2}(\bar{\rho})}{2} + \oint_{\Gamma} \Big[H_{z2}(\bar{\rho}^{'}) \frac{\partial G_{0}(\bar{\rho};\bar{\rho}^{'})}{\partial n^{'}}   +   j\omega \epsilon_{0} \\ E_{\phi2}(\bar{\rho}^{'}) G_{0}(\bar{\rho};\bar{\rho}^{'}) \Big]\ d\Gamma^{'} = H_{z2}^{\mathrm{inc}}(\bar{\rho})\ ; \ \bar{\rho} \in \Gamma
\end{split}
\end{equation}
\end{subequations}
Since the number of unknowns is 8, we need 4 more relations. These are obtained from the metasurface synthesis equations (\ref{CylMSSynthesisEqnsMonoAniso}). From (\ref{CylMSSynthesisEqnsMonoAniso}) we can obtain a matrix relation between $\mathbf{F_{2}(\bar{\rho}^{'})}$ and $\mathbf{F_{1}(\bar{\rho}^{'})}$.
\begin{subequations}
\label{F1F2Eqn}
\begin{align}
\mathbf{F_{2}(\bar{\rho}^{'})} &= \mathbf{A(\bar{\rho}^{'})} \mathbf{F_{1}(\bar{\rho}^{'})} \\
\mathbf{A(\bar{\rho}^{'})} &= \mathbf{A_{2}^{-1}(\bar{\rho}^{'})} \mathbf{A_{1}(\bar{\rho}^{'})}
\end{align}
\end{subequations}
The matrices $\mathbf{A_{1}(\bar{\rho}^{'})}$, $\mathbf{A_{2}(\bar{\rho}^{'})}$ are given by 
\begin{subequations}
\begin{align}
\mathbf{A_{1}(\bar{\rho}^{'})} &= 
\begin{bmatrix}
\frac{j\omega \epsilon_{0} \chi_{\mathrm{ee}}^{\phi z}}{2} & 0 &
\frac{j\omega \epsilon_{0} \chi_{\mathrm{ee}}^{\phi \phi}}{2} & 1 \\
\frac{j\omega \epsilon_{0} \chi_{\mathrm{ee}}^{z z}}{2} & -1 &
\frac{j\omega \epsilon_{0} \chi_{\mathrm{ee}}^{z \phi}}{2} & 0 \\
-1 & \frac{j\omega \mu_{0}\chi_{\mathrm{mm}}^{\phi \phi}}{2} & 0 & \frac{j\omega \mu_{0} \chi_{\mathrm{mm}}^{\phi z}}{2} \\
0 & \frac{j\omega \mu_{0} \chi_{\mathrm{mm}}^{z \phi}}{2} & 1 & \frac{j\omega \mu_{0} \chi_{\mathrm{mm}}^{zz}}{2}
\end{bmatrix} 
\\ 
\mathbf{A_{2}(\bar{\rho}^{'})} &= 
\begin{bmatrix}
\frac{j\omega \epsilon_{0} \chi_{\mathrm{ee}}^{\phi z}}{-2} & 0 &
\frac{j\omega \epsilon_{0} \chi_{\mathrm{ee}}^{\phi \phi}}{-2} & 1 \\
\frac{j\omega \epsilon_{0} \chi_{\mathrm{ee}}^{z z}}{-2} & -1 &
\frac{j\omega \epsilon_{0} \chi_{\mathrm{ee}}^{z \phi}}{-2} & 0 \\
-1 & \frac{j\omega \mu_{0}\chi_{\mathrm{mm}}^{\phi \phi}}{-2} & 0 & \frac{j\omega \mu_{0} \chi_{\mathrm{mm}}^{\phi z}}{-2} \\
0 & \frac{j\omega \mu_{0} \chi_{\mathrm{mm}}^{z \phi}}{-2} & 1 & \frac{j\omega \mu_{0} \chi_{\mathrm{mm}}^{zz}}{-2}
\end{bmatrix} 
\end{align}
\end{subequations}
By using (\ref{F1F2Eqn}), the fields on the inner surface of the metasurface (i.e.$\mathbf{F_{2}}(\bar{\rho}^{'})$) in (\ref{IE2}),(\ref{IE4}) can be replaced with fields on the outer surface of the 	metasurface resulting in the following 2 IEs
\begin{equation}
\label{IE5}
\begin{split}
\oint_{\Gamma} \Big[A_{11}E_{z1}(\bar{\rho}^{'}) +
A_{12}H_{\phi 1}(\bar{\rho}^{'}) + A_{13}E_{\phi 1}(\bar{\rho}^{'}) +
A_{14}H_{z 1}(\bar{\rho}^{'})\Big] \\
 \frac{\partial G_{0}(\bar{\rho};\bar{\rho}^{'})}{\partial n^{'}}   -   j\omega \mu_{0} 
\Big[A_{21}E_{z1}(\bar{\rho}^{'}) + 
A_{22}H_{\phi 1}(\bar{\rho}^{'}) +  A_{23} E_{\phi 1}(\bar{\rho}^{'}) \\ + 
A_{24}H_{z 1}(\bar{\rho}^{'})\Big] G_{0}(\bar{\rho};\bar{\rho}^{'})  d\Gamma^{'}  + 0.5\Big(A_{11} E_{z1}(\bar{\rho}) 
+ A_{12}H_{\phi 1}(\bar{\rho})  \\ +   A_{13}E_{\phi 1}(\bar{\rho}) + A_{14}H_{z 1}(\bar{\rho})\Big) = E_{z2}^{\mathrm{inc}}(\bar{\rho})  \ ; \ \bar{\rho} \in \Gamma
\end{split}
\end{equation}
\begin{equation}
\label{IE6}
\begin{split}
\oint_{\Gamma} \Big[A_{41}E_{z1}(\bar{\rho}^{'}) +
A_{42}H_{\phi 1}(\bar{\rho}^{'}) + A_{43}E_{\phi 1}(\bar{\rho}^{'}) +
A_{44}H_{z 1}(\bar{\rho}^{'})\Big] \\
 \frac{\partial G_{0}(\bar{\rho};\bar{\rho}^{'})}{\partial n^{'}}   +   j\omega \epsilon_{0} 
\Big[A_{31}E_{z1}(\bar{\rho}^{'}) + 
A_{32}H_{\phi 1}(\bar{\rho}^{'}) +  A_{33} E_{\phi 1}(\bar{\rho}^{'}) \\ + 
A_{34}H_{z 1}(\bar{\rho}^{'})\Big] G_{0}(\bar{\rho};\bar{\rho}^{'})  d\Gamma^{'}  + 0.5\Big(A_{41} E_{z1}(\bar{\rho}) 
+ A_{42}H_{\phi 1}(\bar{\rho})  \\ +   A_{43}E_{\phi 1}(\bar{\rho}) + A_{44}H_{z 1}(\bar{\rho})\Big) = H_{z2}^{\mathrm{inc}}(\bar{\rho})  \ ; \ \bar{\rho} \in \Gamma
\end{split}
\end{equation}
The dependence of the elements of matrix $\mathbf{A}$ on $\bar{\rho}^{'}$ is not explicitly shown. The IEs (\ref{IE1}),(\ref{IE5}),(\ref{IE3}),(\ref{IE6}) can be used to solve $\mathbf{F_{1}(\bar{\rho}^{'})}$, which in turn can be substituted in (\ref{F1F2Eqn}) to obtain $\mathbf{F_{2}(\bar{\rho}^{'})}$.
These 4 IEs can be converted to a system of linear equations by using pulse basis function and point matching resulting in the MoM system of equations.
\begin{equation}
\begin{bmatrix}
\mathbf{Z_{11}} & \mathbf{Z_{12}} & \mathbf{Z_{13}} & \mathbf{Z_{14}} \\
\mathbf{Z_{21}} & \mathbf{Z_{22}} & \mathbf{Z_{23}} & \mathbf{Z_{24}} \\
\mathbf{Z_{31}} & \mathbf{Z_{32}} & \mathbf{Z_{33}} & \mathbf{Z_{34}} \\
\mathbf{Z_{41}} & \mathbf{Z_{42}} & \mathbf{Z_{43}} & \mathbf{Z_{44}} 
\end{bmatrix}
\begin{bmatrix}
\mathbf{x_{1}} \\
\mathbf{x_{2}} \\
\mathbf{x_{3}} \\
\mathbf{x_{4}} 
\end{bmatrix}
=
\begin{bmatrix}
\mathbf{b_{1}} \\
\mathbf{b_{2}} \\
\mathbf{b_{3}} \\
\mathbf{b_{4}} 
\end{bmatrix}
\end{equation}
where the unknown vectors are 
\begin{subequations}
\label{unknownveceqn}
\begin{align}
\mathbf{x_{1}} &= [E_{z1,1},\cdots,E_{z1,N}]^{T} \\
\mathbf{x_{2}} &= [H_{\phi 1,1},\cdots,H_{\phi 1,N}]^{T} \\
\mathbf{x_{3}} &= [E_{\phi 1,1},\cdots,E_{\phi 1,N}]^{T} \\
\mathbf{x_{4}} &= [H_{z 1,1},\cdots,H_{z 1,N}]^{T}
\end{align}
\end{subequations}
In (\ref{unknownveceqn}), the number after comma represents the discretization index. Each of the matrices $\mathbf{Z_{ij}}$ are $N \times N$. The matrix elements are as follows
\begin{subequations}
\begin{align}
Z_{11,\mathrm{mn}} &= p_{\mathrm{mn}}\\
Z_{12,\mathrm{mn}} &= r_{\mathrm{mn}} \\
Z_{13,\mathrm{mn}} &= 0 \\
Z_{14,\mathrm{mn}} &= 0 
\end{align}
\end{subequations}
\begin{subequations}
\begin{align}
Z_{21,\mathrm{mn}} &= A_{11,\mathrm{n}}q_{\mathrm{mn}} - A_{21,\mathrm{n}} r_{\mathrm{mn}} \\
Z_{22,\mathrm{mn}} &= A_{12,\mathrm{n}}q_{\mathrm{mn}} - A_{22,\mathrm{n}} r_{\mathrm{mn}} \\
Z_{23,\mathrm{mn}} &= A_{13,\mathrm{n}}q_{\mathrm{mn}} - A_{23,\mathrm{n}} r_{\mathrm{mn}} \\
Z_{24,\mathrm{mn}} &= A_{14,\mathrm{n}}q_{\mathrm{mn}} - A_{24,\mathrm{n}} r_{\mathrm{mn}} 
\end{align}
\end{subequations}
\begin{subequations}
\begin{align}
Z_{31,\mathrm{mn}} &= 0\\
Z_{32,\mathrm{mn}} &= 0 \\
Z_{33,\mathrm{mn}} &= -s_{\mathrm{mn}} \\
Z_{34,\mathrm{mn}} &= p_{\mathrm{mn}} 
\end{align}
\end{subequations}
\begin{subequations}
\begin{align}
Z_{41,\mathrm{mn}} &= A_{41,\mathrm{n}}q_{\mathrm{mn}} + A_{31,\mathrm{n}} s_{\mathrm{mn}} \\
Z_{42,\mathrm{mn}} &= A_{42,\mathrm{n}}q_{\mathrm{mn}} + A_{32,\mathrm{n}} s_{\mathrm{mn}} \\
Z_{43,\mathrm{mn}} &= A_{43,\mathrm{n}}q_{\mathrm{mn}} + A_{33,\mathrm{n}} s_{\mathrm{mn}} \\
Z_{44,\mathrm{mn}} &= A_{44,\mathrm{n}}q_{\mathrm{mn}} + A_{34,\mathrm{n}} s_{\mathrm{mn}} 
\end{align}
\end{subequations}
The coefficients $p_{mn},q_{mn},r_{mn},s_{mn}$ are calculated by
\begin{subequations} 
\begin{align}
p_{\mathrm{mn}} &= \frac{1}{2} \delta_{\mathrm{mn}} - \int_{s_{n}} \! \! \frac{\partial G_{0}(\bar{\rho}_{m};\bar{\rho}^{'})}{\partial n^{'}}\ d\Gamma^{'}  \\
q_{\mathrm{mn}} &= \frac{1}{2} \delta_{\mathrm{mn}} + \int_{s_{n}} \! \! \frac{\partial G_{0}(\bar{\rho}_{m};\bar{\rho}^{'})}{\partial n^{'}}\ d\Gamma^{'}  \\
r_{\mathrm{mn}} &= j\omega \mu_{0} \! \int_{s_{n}} \! \! G_{0}(\bar{\rho}_{m};\bar{\rho}^{'})\ d\Gamma^{'} \\
s_{\mathrm{mn}} &= j\omega \epsilon_{0} \! \int_{s_{n}} \! \! G_{0}(\bar{\rho}_{m};\bar{\rho}^{'})\ d\Gamma^{'} 
\end{align}
\end{subequations}
$s_{n}$ denotes the $n^{\mathrm{th}}$ discretization segment, $\bar{\rho}_{m}$ is the middle point of the $m^{\mathrm{th}}$ segment and $\delta_{mn}$ is the Kronecker delta function. The excitation vector components are given by
\begin{subequations}
\begin{align}
b_{1,m} = E_{z1}^{\mathrm{inc}}(\bar{\rho}_{m})\ , \ b_{2,m} = E_{z2}^{\mathrm{inc}}(\bar{\rho}_{m}) \\
b_{3,m} = H_{z1}^{\mathrm{inc}}(\bar{\rho}_{m})\ , \ b_{4,m} = H_{z2}^{\mathrm{inc}}(\bar{\rho}_{m})
\end{align}
\end{subequations}
Once MoM system of equations are solved to obtain $\mathbf{F}_{1}(\bar{\rho}^{'})$, $\mathbf{F}_{2}(\bar{\rho}^{'})$ are obtained by using (\ref{F1F2Eqn}a).

\section{Numerical validation}
In this section the proposed 2D GSTC-MoM formulation is validated through a case of an anisotropic, gyrotropic circular cylindrical metasurface. For such a metasurface, there are 8 susceptibility components as given in (\ref{CylMSSynthesisEqnsMonoAniso}). To solve for these 8 unknowns (i.e. to synthesize the metasurface), we need two separate field transformations \cite{achouri2015general}. We consider the following field transformations: \textit{Transformation 1}: Field generated by an infinite electric line source, i.e. $\bar{J}_{e} = \delta(x)\delta(y) \hat{z}\ \mathrm{Am^{-2}}$ is transformed to field due to an infinite magnetic line source, i.e. $\bar{J}_{m} = \delta(x)\delta(y) \hat{z}\ \mathrm{Vm^{-2}}$. \textit{Transformation 2}: Field generated by an infinite magnetic line source is attenuated by half.  For both transformations, the metasurface has to be reflection-less. For transformation 1, the electric and magnetic fields on the inner and outer surface	 of the metasurface are given by
\begin{subequations}
\label{Transform1}
\begin{align}
\begin{split}
E_{z}^{(1)-} &= -\frac{\omega \mu_{0}}{4} H_{0}^{(2)}(k_{0}a)\ ; \  H_{\phi}^{(1)-} = -\frac{jk_{0}}{4} H_{1}^{(2)}(k_{0}a) \\   
E_{\phi}^{(1)-} &= 0 \ ; H_{z}^{(1)-} = 0 
\end{split}
\end{align}
\begin{align}
\begin{split}
E_{z}^{(1)+} &= 0 \ ; \  H_{\phi}^{(1)+} = 0 \\   
E_{\phi}^{(1)+} &= \frac{jk_{0}}{4}H_{1}^{(2)}(k_{0}a) \ ; \ H_{z}^{(1)+} = -\frac{\omega \epsilon_{0}}{4}
H_{0}^{(2)}(k_{0}a)
\end{split}
\end{align}
\end{subequations}
Similary for transformation 2, the fields are given by
\begin{subequations}
\label{Transform2}
\begin{align}
\begin{split}
E_{z}^{(2)-} &= 0\ ; \  H_{\phi}^{(2)-} = 0\ \\   
E_{\phi}^{(2)-} &= \frac{jk_{0}}{4}H_{1}^{(2)}(k_{0}a) \ ;\ H_{z}^{(2)-} = -\frac{\omega \epsilon_{0}}{4}H_{0}^{(2)}(k_{0}a) 
\end{split}
\end{align}
\begin{align}
\begin{split}
E_{z}^{(2)+} &= 0 \ ; \  H_{\phi}^{(2)+} = 0 \ \\   
E_{\phi}^{(2)+} &= \frac{jk_{0}}{8}H_{1}^{(2)}(k_{0}a) \ ;\ H_{z}^{(2)+} = -\frac{\omega \epsilon_{0}}{8}
H_{0}^{(2)}(k_{0}a)
\end{split}
\end{align}
\end{subequations}
In (\ref{Transform1}), (\ref{Transform2}), the superscripts $+$ and $-$ denote outer and inner contours respectively, The superscripts $(1)$ and $(2)$ for field components denote  transformation 1 and transformation 2. 
\begin{equation}
\begin{bmatrix}
\mathbf{ZE}^{(1)} & \mathbf{0} \\
\mathbf{0} & \mathbf{ZH}^{(1)} \\
\mathbf{ZE}^{(2)} & \mathbf{0} \\
\mathbf{0} & \mathbf{ZH}^{(2)} 
\end{bmatrix}
\begin{bmatrix}
\chi_{\mathrm{ee}}^{\phi \phi} \\
\chi_{\mathrm{ee}}^{\phi z}    \\
\chi_{\mathrm{ee}}^{z \phi}    \\
\chi_{\mathrm{ee}}^{z z}       \\
\chi_{\mathrm{mm}}^{\phi \phi} \\
\chi_{\mathrm{mm}}^{\phi z}    \\
\chi_{\mathrm{mm}}^{z \phi}    \\
\chi_{\mathrm{mm}}^{z z}       
\end{bmatrix}
= 
\begin{bmatrix}
-\Delta H_{z}^{(1)} \\
\Delta H_{\phi}^{(1)} \\
\Delta E_{z}^{(1)} \\
-\Delta E_{\phi}^{(1)} \\
-\Delta H_{z}^{(2)} \\
\Delta H_{\phi}^{(2)} \\
\Delta E_{z}^{(2)} \\
-\Delta E_{\phi}^{(2)} 
\end{bmatrix}
\end{equation}

where $\mathbf{0}$ is a 2 $\times$ 4 zero matrix, $\mathbf{{ZE}^{(i)}}$, $\mathbf{{ZH}^{(i)}}$ are 
\begin{equation}
\mathbf{{ZE}^{(i)}} = 
j\omega \epsilon_{0} 
\begin{bmatrix}
E_{\phi,\mathrm{av}}^{(i)} &  E_{z, \mathrm{av}}^{(i)}
& 0 & 0 \\
0 & 0 &  E_{\phi,\mathrm{av}}^{(i)} &  E_{z, \mathrm{av}}^{(i)}
\end{bmatrix}
\end{equation}
\begin{equation}
\mathbf{{ZH}^{(i)}} = 
j\omega \mu_{0} 
\begin{bmatrix}
H_{\phi,\mathrm{av}}^{(i)} & H_{z, \mathrm{av}}^{(i)}
& 0 & 0 \\
0 & 0 & H_{\phi,\mathrm{av}}^{(i)} &  H_{z, \mathrm{av}}^{(i)}
\end{bmatrix}
\end{equation}

\begin{figure}[!h]
\centering
\includegraphics[width=3.6in]{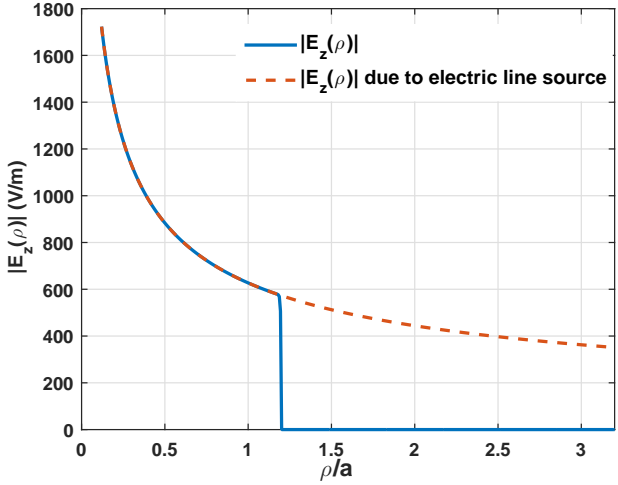}
% where an .eps filename suffix will be assumed under latex, 
% and a .pdf suffix will be assumed for pdflatex; or what has been declared
% via \DeclareGraphicsExtensions.
\caption{Cylindrical metasurface: Magnitude of $E_{z}(\rho)$.}
\label{fig_EzFieldAnisoGyroMS}
\end{figure}

\begin{figure}[!h]
\centering
\includegraphics[width=3.6in]{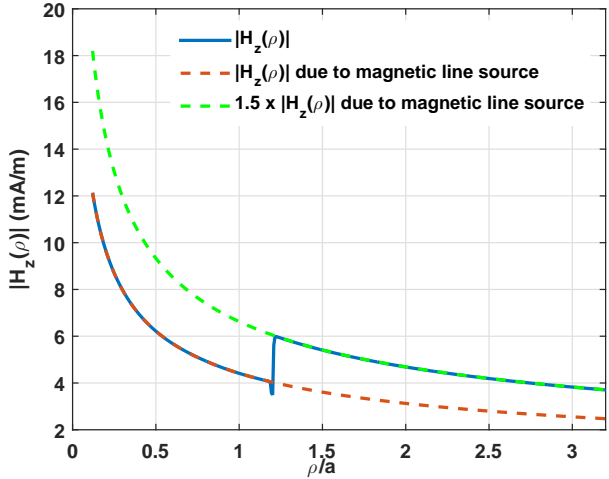}
% where an .eps filename suffix will be assumed under latex, 
% and a .pdf suffix will be assumed for pdflatex; or what has been declared
% via \DeclareGraphicsExtensions.
\caption{Cylindrical metasurface: Magnitude of $H_{z}(\rho)$.}
\label{fig_HzFieldAnisoGyroMS}
\end{figure}

The radius of the cylindrical metasurface  is $a = 1.2 \lambda$. The metasurface is excited simultaneously by both electric and magnetic line source located at $\rho=a$.  This is achieved by setting 
\begin{equation}
\begin{split}
b_{1,m} = 0 \ ,\ b_{2,m} = -\frac{\omega \mu_{0}}{4}H_{0}^{(2)}(k_{0}a) \\
b_{3,m} = 0 \ , \ b_{4,m} = -\frac{\omega \epsilon_{0}}{4}H_{0}^{(2)}(k_{0}a) 
\end{split}
\end{equation}
Therefore the simulation results are expected to be a superposition of both the transformations detailed earlier in this section. The magnitude of the longitudinal fields, $|E_{z}(\rho)|$ and $|H_{z}(\rho)|$ are plotted in Figs. \ref{fig_EzFieldAnisoGyroMS} and \ref{fig_HzFieldAnisoGyroMS}, respectively. Consider the first transformation, i.e. due to electric line source, $\bar{J}_{e} = \delta(x)\delta(y) \hat{z}$. The metasurface was synthesized to be reflection-less for the field generated by electric line source  and to transform the same field into a field due to magnetic line source, $\bar{J}_{m} = \delta(x)\delta(y) \hat{z}$. The reflection-less property can be observed in Fig. \ref{fig_EzFieldAnisoGyroMS}, where the $|E_{z}(\bar{\rho})|$ inside the metasurface (i.e $\rho/a < 1.2$), coincide with $|E_{z}(\bar{\rho})|$ due to an infinite electric line source. $|E_{z}(\bar{\rho})|$ outside the metasurface is zero due to the fact that electric line source field $(\mathrm{TM_{z}}: E{z},H_{\phi})$ is converted to magnetic line source fields $(\mathrm{TE_{z}}:E_{\phi},H_{z})$. Consider the second transformation, i.e. due to magnetic line source $\bar{J}_{m} = \delta(x)\delta(y) \hat{z}$. The metasurface was synthesized to be reflection-less for the field generated by magnetic line source  and to transform the same field by attenuating it by a factor of 2. The reflection-less property can be observed in Fig. \ref{fig_HzFieldAnisoGyroMS}, where the $|H_{z}(\rho)|$ coincide with $|H_{z}(\rho)|$ due to an infinite magnetic line source.  $|H_{z}(\rho)|$ outside the metasurface is sum of the fields due to two transformations. The first transformation results in $|H_{z}(\rho)|$ due to $\bar{J}_{m} = \delta(x)\delta(y) \hat{z}$ and the second transformation results in $|H_{z}(\rho)|$ due to $\bar{J}_{m} = 0.5\delta(x)\delta(y) \hat{z}$. This can be see in Fig. \ref{fig_HzFieldAnisoGyroMS}, where the field outside the metasurface coincide with field due to $\bar{J}_{m} = 1.5\delta(x)\delta(y) \hat{z}$.


\section{Conclusion}
A novel approach based on IE-MoM is provided for fast analysis of circular cylindrical metasurface or circular cylindrical metasurface systems (i.e. layered media separated by cylindrical metasurfaces). The formulation is validated by using an anisotropic, gyrotropic metasurface which can perform two simultaneous field transformations. The formulation can be extended to bianisotropic metasurfaces. In such a case, the matrices $\mathbf{A_{1}}$ and $\mathbf{A_{1}}$ will be more involved. This work also shows the application of bianisotropic susceptibility-GSTC \cite{achouri2015general} approach to cylindrical metasurfaces. For more practical cylindrical metasurface problems, the $\hat{\rho}$ component of the fields cannot be neglected. The GSTC-MoM can be extended to 3D spherical metasurfaces. Future work would include the solution to these two problems, both of which rely on the fundamental principle outlined in this work.

% or


%\appendices

%\section{}

\section*{Acknowledgment}
The first author would like to thank Prof. Jianming Jin of University of Illinois at Urbana-Champaign for his assistance.

\ifCLASSOPTIONcaptionsoff
  \newpage
\fi



\bibliographystyle{IEEEtran}
\bibliography{IEEEabrv,BibDataBase}

\vfill


\end{document}